\documentclass[epj]{svjour}
\usepackage{graphics}
\usepackage{mathrsfs}
\usepackage{amssymb}
\newcommand{\psib}{{\overline{\psi}}}
\newcommand{\chib}{{\overline{\chi}}}
\newcommand{\bfr}{{\mathbf{r}}}
\newcommand{\beq}{\begin{equation}}
\newcommand{\eeq}{\end{equation}}
\newcommand{\beqa}{\begin{eqnarray}}
\newcommand{\eeqa}{\end{eqnarray}}
\newcommand{\mcall}{{\mathbb{L}}}
\newcommand{\mcalb}{{\mathbb{B}}}

\begin{document}
\title{Fermion Bag Approach to Fermion Sign Problems}
\subtitle{New opportunities in lattice fermion field theories}
\author{Shailesh Chandrasekharan\inst{1}% etc
% \thanks is optional - remove next line if not needed
%\thanks{\emph{Present address:} Insert the address here if needed}%
}                     % Do not remove
\offprints{}          % Insert a name or remove this line
\institute{Duke University, Box 90305, Durham NC 27708}
\date{Received: date / Revised version: date}
% The correct dates will be entered by Springer
%
\abstract{
The fermion bag approach is a new method to tackle fermion sign problems in lattice field theories. Using this approach it is possible to solve a class of sign problems that seem unsolvable by traditional methods. The new solutions emerge when partition functions are written in terms of fermion bags and bosonic worldlines. In these new variables it is possible to identify hidden pairing mechanisms which lead to the solutions. The new solutions allow us for the first time to use Monte Carlo methods to solve a variety of interesting lattice field theories, thus creating new opportunities for understanding strongly correlated fermion systems.
 % end of PACS codes
} %end of abstract
\maketitle
\section{Introduction}
\label{intro}

Computing properties of strongly correlated quantum systems, especially with fermions, continues to be an outstanding challenge in quantum many body theory. While Monte Carlo methods can help in such computations, it can be notoriously difficult to design them due to complex phases that arise from quantum interference \cite{Troyer:2004ge}. These complex phases and associated negative signs are essential to produce the novel phenomena found in such systems \cite{Zaanen29022008}. The negative signs make it difficult to identify the correct probability distribution to sample from. For example, if one is interested in computing the thermal properties of a physical system with Hamiltonian $H$ at temperature $T$, the first step in constructing a Monte Carlo algorithm involves rewriting the canonical quantum partition function $Z$ as a classical partition function. This means one writes
\beq
Z = \mathrm{Tr}\Big(\mathrm{e}^{-H/T}\Big)  \ =\ \sum_{C} \ \Omega(C),
\label{qu2cl}
\eeq
where $C$ is an element from a set of classical configurations and $\Omega(C)$ is its Boltzmann weight. The Monte Carlo algorithm is then designed to sample $C$ with probability proportional to $\Omega(C)$. Using Feynman path integrals it is always possible to find classical representations (\ref{qu2cl}) for quantum problems, however in many interesting cases $\Omega(C)$ is either not positive or is not computable in polynomial time. In both cases it becomes difficult to design efficient Monte Carlo algorithms and we say that the classical representation suffers from a sign problem. Solution to this problem involves finding a classical representation with positive Boltzmann weights computable in polynomial time. Finding classical representations of strongly correlated quantum systems free of sign problems is an important field of research at the crossroads of computational and mathematical physics. 

Classical representations of strongly correlated fermion systems are particularly challenging. Due to the Pauli principle, when two identical fermions permute the quantum amplitude picks a negative sign. The spinor components of a fermion can also mix and create further complex Berry phases. The traditional approach for constructing a classical representation exploits the fact that free fermion path integrals can be computed analytically and expressed as the determinant of a fermion matrix. Since fermions can always be envisioned as moving freely in background potentials, in the traditional approach one begins with this view point and integrates over all fermion degrees of freedom uniformly. Thus, the quantum partition function is written as a sum over determinants of matrices that depend on background potentials \cite{Fucito:1980fh,PhysRevLett.46.519}. Since determinants are computable in polynomial time, if they are positive the fermion sign problem is absent and one can design Monte Carlo algorithms \cite{PhysRevLett.47.1628}. However, there are many examples where the determinants can be negative and the sign problem remains unsolved.

The fermion bag approach is an alternative method to deal with fermion path integrals in lattice field theories \cite{Chandrasekharan:2009wc}. Instead of integrating over all fermion degrees of freedom uniformly, the idea is to group them carefully and try to integrate over each group separately. Each group is viewed as a fermion bag and the fermion path integral within each bag is computed analytically. The result of this integration produces the weight of the bag and as long as it is positive, sign problems are absent. The fermion bag approach is an extension of the meron-cluster approach \cite{PhysRevLett.83.3116}, but is applicable more widely.

Fermion bags are not unique and should be defined wisely. Strong and weak coupling expansions can help guide their definition. For example at extremely strong couplings fermions are paired into bosons and fermion sign problems are naturally solved in a proper set of variables. As the coupling becomes weak, free fermions emerge in small regions which can play the role of fermion bags. Similarly when fermions are weakly coupled, Feynman diagrams suggest that fermions be viewed as propagating only between interaction sites which means all interaction sites may be defined as a fermion bag. This diagrammatic approach is well known in condensed matter physics and has become popular recently \cite{PhysRevB.72.035122,RevModPhys.83.349}. There are interesting dualities between strong and weak coupling definitions of fermion bags \cite{PhysRevLett.108.140404}.

The fermion bag approach has a wide variety of applications. In particular when bosonic background fields are written in terms of worldline variables \cite{Chandrasekharan:2008gp}, a new class of sign problems become solvable \cite{PhysRevD.85.091502,PhysRevD.86.021701}. Using a variety of examples, we will discuss some of the opportunities that the new solutions create for understanding strongly correlated fermion systems.

\section{The Fermion Bag Idea}
\label{idea}

The Euclidean quantum partition function for a variety of systems containing fermions can be written schematically as
\beq
Z = \int [d\sigma] \ \mathrm{e}^{-S_b(\sigma)} \ \int [d\psib d\psi] \ \mathrm{e}^{-\psib \ M(\sigma)\ \psi}
\eeq
where $S_b(\sigma)$ is a real bosonic action, $\psib, \psi$ are Grassmann fields representing fermions and $M(\sigma)$ is the fermion matrix that depends on the bosonic fields $\sigma$. In order to construct a classical representation useful for Monte Carlo methods, the traditional approach is to integrate out Grassmann variables and rewrite the partition function as
\beq
Z = \int [d\sigma] \ \mathrm{e}^{-S_b(\sigma)} \ \mathrm{Det}\big(M(\sigma)\big).
\label{trad}
\eeq
If the determinant of the fermion matrix is non-negative the classical representation (\ref{trad}) is free of sign problems.

In many problems where $\mathrm{Det}\big(M(\sigma)\big)$ is non-negative the fermion fields $\psi$ and $\psib$ can be rotated to a basis where the matrix $M(\sigma)$ takes the form
\beq
M(\sigma) = \left(\begin{array}{cc} \ 0 & \ D(\sigma) \cr \ -D^\dagger(\sigma) & \ 0 \end{array}\right),
\label{solvable}
\eeq
which we refer to as the {\em solvable} form. In such cases the partition function can be written as
\beq
Z = \int [d\sigma] \ \mathrm{e}^{-S_b[\sigma]} \ |\mathrm{Det}\big(D(\sigma)\big)|^2,
\label{fpairing}
\eeq
which explicitly shows that the representation of the partition function is free of sign problems. Whenever the fermion matrix has a solvable form (\ref{solvable}) we say that an identifiable pairing mechanism (in a computational sense) exists in the underlying physics. There are interesting problems where such a pairing mechanism is hidden and extra work is necessary to uncover it. For example, in the presence of positive diagonal elements, $\mathrm{Det}\big(M(\sigma)\big)$ is no longer the modulus square of a single determinant, instead it can be written as a weighted sum of modulus squares of determinants of smaller matrices. Unfortunately, there remain many interesting problems where there is no identifiable pairing mechanism and the fermion sign problem seems unsolvable in the traditional approach. As we will see below, the fermion bag approach shows 
that sometimes the pairing mechanism is hidden in background fields and can be recovered by changing the representation of the background fields.

\begin{figure}
\vspace{-0.5in}
\begin{center}
\resizebox{0.45\textwidth}{!}
{\includegraphics{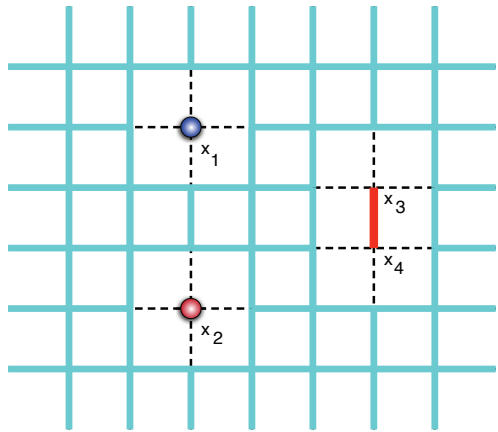} }
\vspace{-0.5in}
\end{center}
\caption{Grassmann integrals can be performed by grouping variables. The above figure illustrates four groups of Grassmann variables discussed in the text. Each group is referred to as a fermion bag.}
\label{fig1}
\end{figure}

The basic idea behind the fermion bag approach is to group Grassmann variables creatively and perform the integration over each group separately. This is in contrast with the traditional approach in which all Grassmann variables are treated uniformly and integrated over. Consider for example the Grassmann integral given by
\beqa
I &=& \int \ [d\psib d\psi] \ \mathrm{e}^{-\psib \ M\ \psi} \ (- U_1\ \psib_{x_1}\psi_{x_1}) (- U_2\ \psib_{x_2}\psi_{x_2}) 
\nonumber \\
&& 
(- U_3 \ \psib_{x_3}\psi_{x_4}\psib_{x_4}\psi_{x_3})
\eeqa
where $M$ is assumed to be in the solvable form (\ref{solvable}). In this example there are four groups of sites that can be treated independently. These are $(x_1)$, $(x_2)$, $(x_3,x_4)$ and all the remaining sites. These four groups are illustrated pictorially in Fig.~\ref{fig1}. We can perform Grassmann integration over each group without worrying about the contribution from the other groups. The integral on  $(x_1)$ gives $U_1$, on $(x_2)$ gives $U_2$ and on $(x_3,x_4)$ gives $U_3$. The remaining Grassmann integral can be also be performed easily by setting the Grassmann variables associated with the three other groups to zero in the exponent. This is equivalent to dropping some rows and the same columns in the matrix $M$ which produces a new matrix $W$. Thus, the above integral becomes
\beq
I = U_1\ U_2\ U_3\ \int \ [d\psib d\psi] \ \mathrm{e}^{-\psib \ W\ \psi} = U_1 U_2 U_3 \mathrm{Det}(W)
\eeq
Clearly, there is no guarantee that $\mathrm{Det}(W)$ is positive. However, if $M$ has the solvable form (\ref{solvable}) as we assumed, then so does $W$ and hence its determinant is also non-negative.  Each group is referred to as a {\em fermion bag}. The three bags $(x_1)$, $(x_2)$ and $(x_3,x_4)$ are referred to as interaction site bags since such terms usually arise from interactions, while the remaining bag is referred to as the free fermion bag since the contribution to it comes from the free fermion action.

In the traditional approach, since all Grassmann variables are treated uniformly, four-fermion couplings have to be converted to fermion bilinears by introducing auxiliary fields. The above discussion suggests that this is not always necessary. In fact sign problems can be introduced through auxiliary fields and may be avoidable using the fermion bag idea. This is illustrated nicely through the following simple plaquette model on a cubic lattice. The partition function of the model is given by
\beq
Z = \int \ [d\psib d\psi] \ \mathrm{e}^{-\psib M \psi + U\sum_{P = wxyz}
\psib_w\psi_w\ \psib_x\psi_x\ \psib_y\psi_y\ \psib_z\psi_z }, 
\eeq
where $\psi_x,\psib_x$ are two Grassmann fields and $P$ represents plaquettes on the lattice. We assume that the matrix $M$ connects only even and odd sites and has the solvable form (\ref{solvable}). For example it can be the free massless staggered fermion matrix. The interaction is a sum over eight-fermion couplings on plaquettes, each term being a product of four $\psib\psi$'s, one from each site of the plaquette. In the traditional approach one could use the identity
\beq
\mathrm{e}^{ U \psib_1\psi_1\psib_2\psi_2\psib_3\psi_3\psib_4\psi_4}
\ =\ \sum_{z \in \mathbb{Z}_4}\ \mathrm{e}^{\big(U^{1/4} \ z \ \sum_{i=1}^4 \psib_i\psi_i\big)}
\eeq
to rewrite the partition function as
\beqa
Z &=& \sum_{[z_P]} \ \int \ [d\psib\psi] \ \mathrm{e}^{-\psib \ \big(M \ +\  U^{1/4} \ \overline{z}\ \big)\psi}
\nonumber \\
&=& \sum_{[z_P]}\ \mathrm{Det}\big(M + U^{1/4} \overline{z}\big)
\eeqa
where $\overline{z}$ is a diagonal matrix with complex elements that depend on the $\mathbb{Z}_4$ plaquette field $[z_P]$. Since $\mathrm{Det}\big(M + U^{1/4} \overline{z}\big)$ can be complex, the traditional approach suffers from a sign problem.  The fermion bag approach on the other hand, is free of sign problems. 

\begin{figure}
\vspace{-0.5in}
\begin{center}
\resizebox{0.45\textwidth}{!}
{\includegraphics{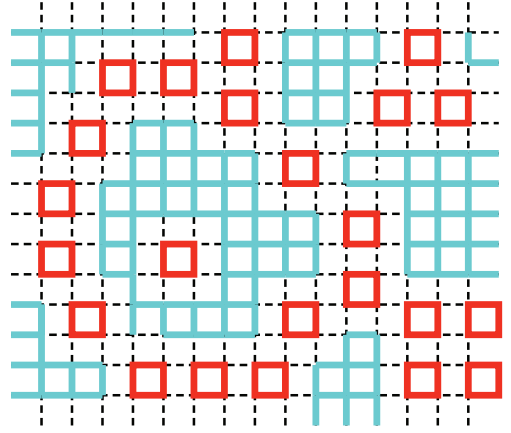} }
\vspace{-0.5in}
\end{center}
\caption{Fermion bags in the plaquette configuration. Each plaquette can be considered as an interacting fermion bag, while the remaining sites as a free fermion bag. At large couplings the free fermion bag can split into many smaller bags.}
\label{fig:2}
\end{figure}

Instead of introducing auxiliary fields, in the fermion bag approach we can expand the interaction term as
\beq
\mathrm{e}^{U \prod_{x\in P} \psib_x\psi_x } = \sum_{n_P = 0,1}\ 
\big( U \prod_{x\in P} \psib_x\psi_x \big)^{n_P}.
\eeq
Introducing a discrete plaquette field $n_P = 0,1$ it is possible to write the above partition function as
\beqa
Z &=& \sum_{[n_P]} \ \int \ [d\psib\psi] \ \mathrm{e}^{-\psib \ M\ \psi} \prod_{P,\ x\in P}
\big( U \prod_{x\in P} \psib_x\psi_x \big)^{n_P}.
\eeqa
An illustration of the plaquette field on a square lattice is shown in Fig.~\ref{fig:2}. We can now group Grassmann variables associated with each plaquette where $n_p=1$ into separate bags and the remaining sites which do not touch any plaquettes into a separate bag. Performing the integrals separately the fermion bag partition function takes the form
\beq
Z \ =\ \sum_{[n_P]}\ U^{N_P}\ \mathrm{Det}\big(W(n_P)\big),
\eeq
where $N_P$ is the total number of $n_p=1$ plaquettes in the configuration $[n_P]$. The matrix $W(n_P)$ is obtained from the matrix $M$ by dropping the rows and columns associated with sites that belong to all non-zero plaquettes. Since $W(n_P)$ also has the solvable form (\ref{solvable}), $\mathrm{Det}\big(W[n]\big) \geq 0$ and the sign problem is absent.

\section{New Class of Solvable Models}
\label{sec3}

While the plaquette model discussed above provides a simple example to demonstrate the usefulness of the fermion bag idea, the full advantage of the method to solve sign problems emerges in the presence of dynamical bosonic fields. In this section we explain how a new class of lattice field theories which contain sign problems when formulated in the traditional approach, can be written without sign problems in the fermion bag approach. The pairing mechanism, which seems absent in the traditional method, can be resurrected by rewriting some bosonic fields in worldline variables. These new solvable models are described by the following general action:
\beq
S = \sum_{x,y} \psib_x \big(M(\sigma) \ +\ g\ \Phi(\phi)\big)_{xy} \psi_y + S_b(\sigma,\phi),
\label{newclass}
\eeq
where the index $x=1,2,..,V$ labels both space-time and other internal indices, $\psi_x,\psib_x$ are Grassmann fields, $M(\sigma)$ has the solvable form (\ref{solvable}) and $(\Phi)_{xy} = \phi_{x} \delta_{x,y}$ is a diagonal matrix that depends on an additional complex scalar field $\phi_x$. Importantly, $M(\sigma)$ does not depend on $\phi$ and the coupling $g$ is real and positive. The presence of an arbitrary complex diagonal matrix $g\Phi(\phi)$ makes $\mathrm{Det}(M(\sigma)+g \Phi(\phi))$ complex, implying that there is no pairing mechanism in the traditional approach and it suffers from a sign problem irrespective of the form of the bosonic action. However, this sign problem is completely solvable in the fermion bag approach if the bosonic action $S_b(\sigma,\phi)$ is constrained to have the {\em solvable} form as explained below.

The bosonic action $S_b(\sigma,\phi)$ will be defined as {\em solvable} if the $k$-point correlation function
\beq
G(x_1,...,x_k,\sigma)\ =\ \int [d\phi] \mathrm{e}^{-S_b(\sigma,\phi)}\ \phi_{x_1}\ ...\ \phi_{x_k}
\label{boscorr1}
\eeq
can be expressed as a sum over positive Boltzmann weights computable in polynomial time. In the next section we will argue that many physically interesting bosonic actions are solvable. Writing $\phi = \rho\ \mathrm{e}^{i\theta}$ we will show that for these theories we can write
\beq
G(x_1,...,x_k,\sigma)\ = \ \sum_{[b]} \int \ [d\rho]\ \Omega(b,\sigma,\rho,n),
\label{boscorr2}
\eeq
where $[b]$ is a worldline configuration of charged particles described by integer valued bond variables on the lattice, $[n]$ is a configuration of monomers associated to the $k$ sites $x_1,x_2,...,x_k$ and $\Omega \geq 0$ is the Boltzmann weight calculable in polynomial time. Such worldline representations for bosonic field theories have become popular recently \cite{Chandrasekharan:2008gp,PhysRevD.81.125007,Wolff:2009kp,Gattringer:2012df}.  

Assuming that $S_b(\sigma,\phi)$ is solvable let us now prove that the class of lattice field theories with action (\ref{newclass}) can be written without sign problems in the fermion bag approach. We expand the interaction term at each lattice site $x$ as
\beq
\mathrm{e}^{-g \phi_x \psib_x\psi_x}\ =\ \sum_{n_x = 0,1} (-g \phi_x \psib_x\psi_x)^{n_x}
\eeq
where $n_x=1 (0)$ is a monomer field that indicates the presence (or absence) of a monomer. After introducing the monomer field $[n]$, the partition function takes the form
\beqa
Z &=& \sum_{[n]} \int [d\sigma] [d\phi] \ \mathrm{e}^{-S_b(\phi,\sigma)}
\nonumber \\
&& \int [d\psib\ d\psi]\ \mathrm{e}^{- \psib M(\sigma)\psi} \prod_{x} (-g \phi_x \psib_x\psi_x)^{n_x}
\eeqa
Assuming that $x_1,x_2,...,x_k$ are the sites where $n_x = 1$, the partition function can be written as
\beqa
&& Z \ =\ \int [d\sigma] \ \sum_{[n]} g^k \Big\{\int\ [d\phi] \ 
\mathrm{e}^{-S_b(\phi,\sigma)}\ \phi_{x_1} \ ...\ \phi_{x_k}\Big\} \times
\nonumber \\
&& 
\Big\{ \int [d\psib\ d\psi]\ \mathrm{e}^{- \psib M(\sigma)\psi}  
(-\psib_{x_1}\psi_{x_1})...(-\psib_{x_k}\psi_{x_k})\Big\}.
\label{partfn2}
\eeqa
The term within the first curly braket is the $k$-point correlation function defined in (\ref{boscorr1}). Let us assume that we can use (\ref{boscorr2}) to express it as
\beq
\int\ [d\phi] \ 
\mathrm{e}^{-S_b(\phi,\sigma)}\ \phi_{x_1} \ ...\ \phi_{x_k} =
\sum_{[b]} \int \ [d\rho]\ \Omega(b,\sigma,\rho,n).
\label{boscorr3}
\eeq
The second curly braket is the $k$-point fermion correlation function and we use the fermion bag idea to compute it. We divide the Grassmann integral into $k+1$ groups, one for each of the $k$ interaction sites and one group that includes all the remaining free sites. The integral over each of the $k$ sites is trivial and gives the identity. The remaining Grassmann integral is simply the determinant of the $(V-k)\times (V-k)$ matrix $W(\sigma,n)$ obtained from the matrix $M(\sigma)$ by dropping the rows and columns that belong to the $k$ monomer sites. If $M(\sigma)$ has the solvable form (\ref{solvable}) so does $W(\sigma,n)$, which means
$\mathrm{Det}\big(W(\sigma,n)\big) \geq 0$. Hence the $k$-point fermion correlation function is given by
\beqa
\int [d\psib\ d\psi]\ &\mathrm{e}^{- \psib M[\sigma]\psi}& 
(-\psib_{x_1}\psi_{x_1})...(-\psib_{x_k}\psi_{x_k})
\nonumber \\
&=& \mathrm{Det}\big(W(\sigma,n)\big).
\label{fermcorr}
\eeqa
Combining (\ref{partfn2}), (\ref{boscorr3}), and (\ref{fermcorr}) we obtain
\beq
Z = \int [d\sigma d\rho] \ \sum_{[n,b]} \ g^k \ \Omega(b,\sigma,\rho,n) \ 
\mathrm{Det}(W(\sigma,n)),
\eeq
which is free of sign problems. 

Instead of using the fermion bag idea we can also use Wick's theorem to compute the correlation function in (\ref{fermcorr}). This leads to a sum over Feynman Diagrams which can be performed exactly. One obtains
\beqa
& \Big\{ \int [d\psib\ d\psi]\ &\mathrm{e}^{- \psib M[\sigma]\psi}  
(-\psib_{x_1}\psi_{x_1})...(-\psib_{x_k}\psi_{x_k})\Big\} 
\nonumber \\
&& = \ \ \  \mathrm{Det}\big(M(\sigma)\big)\ \mathrm{Det}\big(G(\sigma,n)\big),
\label{fermcorr2}
\eeqa
where $G(\sigma,n)$ is the $k \times k$ propagator matrix connecting the $k$ monomer sites $x_1,...x_k$. This diagrammatic approach has been exploited in condensed matter physics \cite{PhysRevB.72.035122,RevModPhys.83.349}. Combining (\ref{fermcorr}) and (\ref{fermcorr2}), we can derive the duality relation
\beq
\mathrm{Det}\big(W(\sigma)\big) = \mathrm{Det}\big(M(\sigma)\big)\ \mathrm{Det}\big(G(\sigma,n)\big).
\eeq
which connects the diagrammatic approach to the fermion bag idea \cite{PhysRevLett.108.140404}. From a computational point of view, when $k$ is large the left hand side is easier to calculate and when $k$ is small the right hand side becomes easier.

\section{Solvable Bosonic Actions}
\label{sec4}

In the previous section we defined a bosonic action $S_b(\sigma,\phi)$ to be {\em solvable} if the $k$-point correlation function of the complex field $\phi$ could be expressed as a sum of positive Boltzmann weights computable in polynomial time. In this section we argue that many standard actions are solvable in this sense.

Assuming $\phi = \rho \ \mathrm{e}^{i\theta}$ let us first argue that the general bosonic action of the form
\beq
S_b(\sigma,\phi) =  s(\sigma,\rho) - \sum_{x,y} \beta_{xy}\ \cos(\varepsilon_x\theta_x + \varepsilon_y\theta_y)
\label{boseclass}
\eeq
is solvable if $s(\sigma,\rho)$ is a real function and $\beta_{xy} \geq 0$ are non-negative real numbers for all values of $x$ and $y$. $\varepsilon_x = \pm 1$ is an arbitrary Ising field whose precise form is not relevant for the moment. Substituting the above action in (\ref{boscorr1}) we obtain
\beqa
&& G(x_1,...,x_k,\sigma)\ = \ \int \ [d\rho]\ \mathrm{e}^{-s(\sigma,\rho)} \rho_{x_1} ...\rho_{x_k}\ 
\nonumber \\
&&\ \ \times \int [d\theta] \ \prod_{(x,y)}\ 
\mathrm{e}^{\beta_{xy} \cos(\varepsilon_x\theta_x + \varepsilon_y\theta_y)}
\ \ \mathrm{e}^{i\theta_{x_1}}\ ...\ \mathrm{e}^{i\theta_{x_k}}
\eeqa
Using the identity\footnote{$I_b(\beta)$ is the modified Bessel function}
\beq
\mathrm{e}^{\beta_{xy} \cos(\varepsilon_x\theta_x + \varepsilon_y\theta_y)}
\ =\ 
\sum_{b_{xy}=-\infty}^\infty \ I_{b_{xy}} (\beta_{xy}) \ \mathrm{e}^{i b_{xy} (\varepsilon_x\theta_x+\varepsilon_y\theta_y)}
\eeq
on each bond $(x,y)$ and identifying $b_{xy}$ as an integer current variable on bonds we obtain
\beqa
&& G(x_1,...,x_k,\sigma)\ = \ \sum_{[b]} \int \ [d\rho]\ \mathrm{e}^{-s(\sigma,\rho)} \rho_{x_1} ...\rho_{x_k}
\nonumber \\
&& \times \ 
\Big(\prod_{(xy)}\ I_{b_{xy}} (\beta_{xy})\Big)\ \ 
\Big(\prod_{x} \ \delta(n_x + \sum_{y} \varepsilon_x (b_{xy} + b_{yx})\Big).
\nonumber \\
\label{boscorr4}
\eeqa
where we have defined $b$ as the current (or worldline) configuration, $n$ as a monomer field such that $n_x = 1$ at the $k$ sites $x_1,x_2,..,x_k$ and $n_x = 0$ on other sites, and  computed the $\theta$ integral.
Comparing with (\ref{boscorr2}) we can identify
\beqa
\Omega(b,\sigma,\rho,n) &=& \mathrm{e}^{-s(\sigma,\rho)}\ \rho_{x_1}\ ...\ \rho_{x_k}\ 
 \Big(\prod_{(xy)}\ I_{b_{xy}} (\beta_{xy})\Big)
\nonumber \\
&& \times \ 
\Big(\prod_{x} \ \delta(n_x + \sum_{y} \varepsilon_x (b_{xy} + b_{yx})\Big).
\eeqa
Since $\Omega(n,\sigma,\rho,n)$ is non-negative and computable in polynomial time, the bosonic action (\ref{boseclass}) is indeed solvable. 

Many standard bosonic actions can be expressed in the form of (\ref{boseclass}). As a first example, consider the standard complex scalar field theory with quartic coupling on a lattice whose action is
\beq
S(\phi) = \sum_{x} (\kappa |\phi_x|^2 + \lambda |\phi_x|^4)
-\beta \sum_{x,\alpha} (\phi_x \phi^*_{x+\hat{\alpha}} + \phi_{x+\hat{\alpha}} \phi^*_x)
\label{standardbose}
\eeq
where $x$ is a lattice site on a hyper-cubic lattice and $\hat{\alpha}$ is the unit vector along each direction. The standard choice also implies $\beta,\lambda \geq 0$. Correlation functions of $\phi$ determine the solvability of the action. Expressing the complex field in polar form ($\phi=\rho \mathrm{e}^{i\theta})$) we obtain
\beq
S(\phi) = \sum_{x} (\kappa \rho_x^2 + \lambda \rho_x^4) -
\sum_{x,\alpha} \beta \rho_x\rho_{x+\hat{\alpha}} \cos(\theta_x - \theta_{x+\hat{\alpha}})
\label{sbose}
\eeq
which is indeed of the form (\ref{boseclass}).

As a second example we consider an $O(4)$ invariant scalar field theory involving four real fields denoted as $\vec{\varphi} = (\sigma,\vec{\pi})$ whose lattice action is given by
\beq
S_b(\vec{\varphi}) = \sum_x \Big\{ \kappa (\vec{\varphi}_{x} \cdot \vec{\varphi}_{x}) + 
\lambda (\vec{\varphi}_{x} \cdot \vec{\varphi}_{x})^2 \Big\} - 
\beta \sum_{x,\alpha} \vec{\varphi}_x \cdot \vec{\varphi}_{x+\hat{\alpha}},
\label{o4model}
\eeq
For reasons that will become clear later, we choose to identify $\phi = -i(\pi_1 - \pi_2)$ as the complex scalar field whose correlation functions determine the solvability of the model. Assuming $\phi = \rho \mathrm{e}^{i\theta}$ we can rewrite the action as
\beq
S_b(\sigma,\pi_3,\phi) =  s(\sigma,\pi_3,\rho) -\sum_{x,\alpha} \beta \rho_x\rho_{x+\alpha}\cos(\theta_x-\theta_{x+\alpha})
\label{o4model1}
\eeq
where
\beqa
s(\sigma,\pi_3,\rho) &=& \sum_{x} \big( \kappa(\sigma_x^2 + \pi_3^2 + \rho_x^2)  + 
\lambda (\sigma_x^2 + \pi_3^2 + \rho_x^2)^2 \big)
\nonumber \\
&-& \beta \sum_{x,\alpha} \big(\sigma_x\sigma_{x+\alpha} + \pi_{3,x} \pi_{3,x+\hat{\alpha}}\big) 
\eeqa
Again clearly (\ref{o4model1}) is of the form (\ref{boseclass}).

As a final example consider a non-relativistic scalar field theory described by two complex scalar fields $\varphi_1(\bfr,t)$ and $\varphi_2(\bfr,t)$ where $\bfr$ labels the spatial coordinates and $t$ the temporal coordinate. The lattice action of interest is given by
\beqa
&& S_b(\varphi_1,\varphi_2) = \sum_{\bfr,t} \kappa_1 |\varphi_1(\bfr,t)|^2 + \kappa_2 |\varphi_2(\bfr,t)|^2) 
\nonumber \\
&& - \sum_{\langle\bfr,\bfr'\rangle,i,j} \beta_{s,ij}
\big(\varphi_i(\bfr,t)\varphi_j^*(\bfr',t) + \varphi_i^*(\bfr,t)\varphi_j(\bfr',t)\big)
\nonumber \\
&& - \sum_{\langle t,t'\rangle,\bfr} \beta_{t,ij}
\big(\varphi_i(\bfr,t)\varphi_j^*(\bfr,t') + \varphi_i^*(\bfr,t)\varphi_j(\bfr,t')\big),
\label{bosepairing}
\eeqa
where $\langle \bfr,\bfr'\rangle$ and $\langle t t'\rangle$ denote the nearest neighbor sites in space and time respectively. In this model, arbitrary correlation functions of both complex fields $\varphi_1$ and $\varphi_2$ determine the solvability of the action. We can combine both fields into a single complex field $\phi_x$ by defining that the index $x$ spans space-time lattice points two times. Assuming $V$ is the total number of space-time points, we can define $\phi_x = \varphi_1(\bfr,t)$ while $\phi_{x+V} = \varphi_2(\bfr,t)$. Then correlation functions of $\phi_x$ will determine the solvability of the model. Assuming $\phi_x=\rho_x \mathrm{e}^{i\theta_x}$ the action takes the solvable form (\ref{boseclass})
\beq
S_b(\phi) = \sum_{x} \kappa_x \rho_x^2 - 
\sum_{x,y} \beta_{xy}  \rho_x \rho_y \cos(\theta_x - \theta_y).
\label{bosepairing1}
\eeq
if $\kappa_x$ and $\beta_{xy}$ are defined appropriately in terms of $\kappa_1$, $\kappa_2$, $\beta_{s,ij}$ and $\beta_{t,ij}$.

\section{New Opportunities}
\label{sec5}

The new solutions to fermion sign problems discussed in Sect.~\ref{sec3} create new opportunities to understand strongly correlated fermion systems. In this section we construct three classes of models in different areas of physics that were intractable earlier due to sign problems, but thanks to the new solutions they can now be used to address interesting questions.

\subsection{Quantum Criticality with Staggered Fermions}

It is well known that massless fermions can become massive due to strong interactions. Quantum critical points often separate massive and massless phases. In $3+1$ dimensions, critical points are rare and gauge fields play an important role in creating them \cite{Appelquist:1996dq}. On the other hand in $2+1$ dimensions many critical points are known to exist even in the absence of gauge fields \cite{Rosenstein:1990nm}. In either case, understanding their properties is an exciting field of research with applications both in particle physics \cite{Giedt:2012it,Neil:2012cb} and condensed matter physics \cite{PhysRevLett.97.146401}. Since these critical points do not occur in a perturbative regime, Monte Carlo methods are essential to compute their properties. 

Properties of critical points in $2+1$ dimensions can be studied either through four-fermion models or Yukawa models \cite{Hands:1992be,AnnPhys.224.29}. Effects of long range interactions and disorder are also interesting to explore \cite{PhysRevB.75.235423,PhysRevB.77.075115,PhysRevB.80.075432,PhysRevB.80.081405}. Unfortunately, traditional Monte Carlo methods can only be used in a small subset of these models where sign problems are solvable \cite{Karkkainen:1993ef,Hands:1995jq,PhysRevD.53.4616,DelDebbio:1995zc,Debbio:1997dv,Barbour:1998yc,Christofi:2006zt,PhysRevB.79.241405,PhysRevLett.102.026802,PhysRevB.81.125105,PhysRevB.72.085123,nature.464.08942,Sorella2012}. A large number of equally interesting models have remained unexplored due to sign problems. Fortunately, some of these can now be studied using the fermion bag approach.

Consider $N$-flavors of staggered fermions denoted by Grassmann fields $\chib_{x,i}, \chi_{x,i}, i=1,2,..,N$, interacting with a complex scalar field $\phi_x = \rho_x\mathrm{e}^{i\theta_x}$ through a Yukawa coupling in three space-time dimensions. The lattice action is given by
\beq
S = S_b(\phi) + \sum_{x,y,i} \chib_{x,i} D^{(s)}_{xy} \chi_{y,i} \ +\ g\ \sum_{x,i} \rho_x \ 
\mathrm{e}^{i\varepsilon_x \theta_x} \ \chib_{x,i} \chi_{x,i} 
\label{3dstagg}
\eeq
where $x,y$ denote the sites on a cubic lattice, $\langle x y \rangle$ denotes nearest neighbor sites, $\varepsilon_x = (-1)^{x_1+x_2+x_3}$ assuming $x_1,x_2,x_3$ denote the three coordinates of the lattice site $x$. We assume the coupling $g$ to be real and positive. The massless lattice staggered Dirac matrix $D^{(s)}_{xy}$ is given by
\beq
D^{(s)}_{xy} = \frac{1}{2}\sum_{\alpha=1,2,3} \eta_{x,\alpha} \ 
\Big\{ \delta_{x+\hat{\alpha,}y} - \delta_{x-\hat{\alpha},y} \Big\}
\label{3dstaggdirac}
\eeq
with $\eta_{x,1} =1, \eta_{x,2} = (-1)^{x_1}, \eta_{x,3} = (-1)^{x_1+x_2}$ denoting the staggered fermion phases. Using the properties of the staggered Dirac operator and by denoting the complex field on even sites as $\varphi_e$ and those on the odd sites as $\varphi_o$, in an appropriate basis ({\ref{3dstagg}) can be written as (\ref{newclass}) where
\beq
M + g \ \Phi(\phi) = \left(\begin{array}{ccc} \  g \varphi_e  && D \cr\cr
-D^\dagger  &&  g \varphi_o^* \end{array}\right).\ \ 
\eeq
Here $D$ is the sub-matrix of the full matrix $D^{(s)}$ that connects odd sites with even sites. Since $\mathrm{Det}\big(M + g\ \Phi\big)$ can be complex in general, the models described by (\ref{3dstagg}) suffer from sign problems in the traditional approach. However, if the bosonic action is solvable and for example takes the form (see (\ref{boseclass}))
\beq
S(\phi) = \sum_{x} (\kappa \rho_x^2 + \lambda \rho_x^4) -
\sum_{x,y} \beta_{xy} \rho_x\rho_{y} \cos(\varepsilon_x\theta_x + \varepsilon_y\theta_y),
\eeq
these models are free of sign problems in the fermion bag approach.

Models described by (\ref{3dstagg}) possess a rich phase diagram especially when the freedom in the bosonic action is exploited. There are massless and massive fermion phases and the critical points that separate the two have properties that can depend on $N$ and the symmetry that is broken. By choosing the bosonic action carefully we can change the symmetries of the model. For example if $\beta_{xy}$ is non-zero only between even and odd sites, then it is easy to verify that the action (\ref{3dstagg}) is invariant under the $U(1)$ chiral symmetry 
\beq
\psi_{x,i} \rightarrow \mathrm{e}^{i\varepsilon_x\varphi/2} \psi_{x,i}, \ \ 
\psib_{x,i} \rightarrow \mathrm{e}^{i\varepsilon_x\varphi/2} \psib_{x,i},\  
\mathrm{e}^{i\theta_x} \rightarrow \mathrm{e}^{i(\theta_x-\varphi)}.
\eeq
One the other hand if we allow $\beta_{xy}$ to be arbitrary,  the action (\ref{3dstagg}) is invariant only under a $Z_2$ symmetry
\beq
\psi_{x,i} \rightarrow \mathrm{e}^{i\varepsilon_x\pi/2} \psi_{x,i}, \ \ 
\psib_{x,i} \rightarrow \mathrm{e}^{i\varepsilon_x\pi/2} \psib_{x,i},\  
\mathrm{e}^{i\theta_x} \rightarrow \mathrm{e}^{i(\theta_x-\pi)}.
\eeq
It is also easy to introduce various kinds of disorder and long range interactions through the bosonic action and the Yukawa couplings.

\subsection{Pairing Interactions with Hamiltonian Fermions}

The idea of pairing plays a central role in our understanding of superfluidity and superconductivity. It has a variety of applications from materials physics \cite{RevModPhys.84.1383} to neutron stars \cite{RevModPhys.75.607}. Many types of pairings have been discovered in nature. For example spin-singlet s-wave pairing is the simplest and leads to conventional superconductivity. On the other hand spin-triplet pairing is also well known and leads to unconventional superconductivity with many interesting properties \cite{RevModPhys.75.657}. The well known high $T_c$ materials contain d-wave pairing and continue to be interesting even today \cite{RevModPhys.81.481}. 

From a computational point of view, pairing also plays an important role in solutions to sign problems. As we already pointed out in Sect.~\ref{idea}, if the fermion matrix has the solvable form (\ref{solvable}) then a pairing mechanism (in a computational sense) exists in the model. In a dense fermion system such pairing mechanisms can easily be lost especially when fermions interact with bosonic fields. While QCD at finite baryon densities is a famous example of this feature of dense fermion systems, many simpler models exhibit similar features. The fermion bag approach is able to uncover hidden pairing mechanisms and hence is able solve new sign problems. Below we construct the simplest class of such models with hidden s-wave pairing. With some effort it should be possible to extend these ideas to pairings in other channels. In order to demonstrate the wide applicability of our ideas, here we choose to work with Hamiltonian lattice fermions.

Consider the free Hamiltonian of a spin-less fermion given by
\beq
H = \sum_{\bfr',\bfr} c^\dagger_{\bfr'} \ H_{\bfr',\bfr} \ c_\bfr
\eeq
where $c_\bfr$ and $c^\dagger_\bfr$ are fermion annihilation and creation operators at the lattice site $\bfr$ and $H_{\bfr',\bfr}$ is the one-particle operator that describes fermion hopping on the lattice. It is well known (see \cite{Lee:2008fa} for a recent review) that the corresponding partition function can be written as a Grassmann integral on a space-time lattice, whose action is
\beq
S = \sum_{t',\bfr',t,\bfr} \ \chib_{t',\bfr'} \ D_{t',\bfr';t,\bfr} \ \chi_{t,\bfr}
\eeq
where $D$ is the non-relativistic matrix given by
\beq
D_{t',\bfr';t,\bfr} = -\delta_{t'+1,t}\delta_{\bfr',\bfr} + 
\delta_{t',t}\big(\delta_{\bfr',\bfr} - \varepsilon H_{\bfr',\bfr}\big).
\eeq
Assuming $L_t$ time slices, the temporal continuum limit is obtained by taking $\varepsilon \rightarrow 0$ and $L_t \rightarrow \infty$ while keeping $\varepsilon L_t = 1/T$ (the inverse temperature) fixed. Finite densities can be introduced with a chemical potential that enters through the Hamiltonian \cite{Lee:2008fa}.

With this notation consider two species of fermions labeled with Grassmann fields $\chib_{t,\bfr,i}, \chi_{t,\bfr,i},i=1,2$ which interact with two complex scalar fields $\varphi_1$ and $\varphi_2$. The action of is given by
\beqa
&& S \ =\  S_b(\varphi_1,\varphi_2)  +\  \sum_{t',\bfr',t,\bfr,i} \chib_{t',\bfr',i}\ D_{t',\bfr';t,\bfr} \ \chi_{t,\bfr,i}
\nonumber \\
&& \!\!\!\!\!\!\!\! + \ g \! \sum_{t',\bfr',t,\bfr,i} \!\!
\Big(\varphi_2^*(\bfr,t) \chi_{t,\bfr,1}\chi_{t,\bfr,2} + \varphi_1(\bfr,t) \chib_{t,\bfr,2}\chib_{t,\bfr,1}\Big)
\label{fpairact}
\eeqa
where $S_b(\varphi_1,\varphi_2)$ will be assumed to be solvable and could take the form (\ref{bosepairing}). The action is invariant under $SU(2)$ transformations 
\beq
\chi_{t,\bfr,i} \rightarrow \sum_j\ U_{ij} \chi_{t,\bfr,j},\ \ \chib_{t,\bfr,i} \rightarrow \sum_j\ 
\chib_{t,\bfr,j} {U^\dagger}_{ji} 
\eeq
where $U \in SU(2)$ and $U(1)$ fermion number transformation 
\beqa
&\chi_{t,\bfr,i} \rightarrow \mathrm{e}^{i\theta}\ \chi_{t,\bfr,i}, \ \ \ 
\chib_{t,\bfr,i} \rightarrow \chib_{t,\bfr,i} \ \mathrm{e}^{-i\theta},\ &
\nonumber \\
& \varphi_i(\bfr,t) \rightarrow \varphi_i(\bfr,t) \ \mathrm{e}^{i2\theta}. &
\eeqa
Although the action (\ref{fpairact}) seems to contain pairing interactions the complex scalar field hides it. Indeed the fermion determinant in a fixed scalar field background can be complex and the models are intractable in the traditional approach due to sign problems. By rewriting the bosonic variables in the world line representation one can resurrect the lost pairing mechanism in the fermion bag approach. This solves the sign problem.

To see that (\ref{fpairact}) can be written as (\ref{newclass}) we perform the following transformations
\beqa
&\chib_{t,\bfr,1} \rightarrow -\psib_{t,\bfr,1},\ \chi_{t,\bfr,1} \rightarrow -\psi_{t,\bfr,2},&
\nonumber \\
&\chib_{t,\bfr,2} \rightarrow \psi_{t,\bfr,1},\ \chi_{t,\bfr,2} \rightarrow \psib_{t,\bfr,2}.\ &
\eeqa
In addition using the fact that $D^T = D^\dagger$, we can verify that (\ref{fpairact}) indeed takes the solvable form (\ref{newclass})  where in the $2\times 2$ space of the fermion species we get
\beq
M + g \ \Phi = \left(\begin{array}{ccc} \  g\varphi_1  && D \cr\cr
-D^\dagger  &&  g\varphi_2^* \end{array}\right).\ \ 
\eeq
Since $S_b(\varphi_1,\varphi)2)$ is solvable, the partition function can be expressed without sign problems in the fermion bag approach.

Traditionally, the above class of models have been studied only in the limit where the bosonic fields behave like auxiliary fields. When they are integrated out one obtains the attractive Hubbard model, which can be formulated without sign problems. However, when bosons remain dynamical traditional methods cannot solve the associated sign problems. Fortunately, since the fermion bag approach is free of sign problems, one can use these models to understand the physics of pairing in more generality. By exploiting the freedom in the choice of bosonic actions, one should be able to explore many new regimes and scales that have remained unexplored. Further, as we explain in Sect.~\ref{sec6} below, it should be possible to make $H_{\bfr',\bfr}$ more interesting by introducing spin-orbit couplings and thus allowing us to extend the limits of possibility even further.

\subsection{Yukawa Models with Wilson Fermions}

Long ago in his pioneering work \cite{Yukawa:1935xg} Yukawa suggested that nuclear physics may be governed by a theory of massive nucleons interacting through the exchange of pions. Today we understand that this approach to nuclear physics can be made more systematic through the use of nuclear effective field theory (NEFT) \cite{Bedaque:2002mn,Epelbaum:2008ga}. In fact lattice formulations of NEFT have become popular recently \cite{Lee:2004si} and some properties of low lying nuclei have also been computed \cite{Epelbaum:2009pd}. From a more conceptual point of view, while NEFT without pions is on a strong footing, there is disagreement on how to incorporate the dynamics of pions systematically \cite{Fleming2000,Beane:2001bc,Harada:2010ba}. Symmetries should ultimately determine how to incorporate them and it is possible that a non-perturbative framework is necessary.

It is interesting to explore if lattice field theory models that contain the same symmetries and low energy degrees of freedom as NEFT, can help us understand how to incorporate the dynamics of pions. Lattice Yukawa models are ideal starting grounds for such an exploration \cite{deSoto:2011sy}. For example, in the simplest setting one could start with two flavors of light Dirac fermions coupled to an $O(4)$ scalar field through an $SU(2)_L \times SU(2)_R \equiv O(4)$ symmetric Yukawa coupling. In a phase where the $O(4)$ symmetry breaks spontaneously to $O(3)$, the low energy physics would contain two massive fermions and three massless Goldstone bosons which are precisely the degrees of freedom of NEFT. Further, the couplings of the fermions to the bosons would be correctly constrained by the relevant symmetries \cite{Chandrasekharan:2003wy}. Since it is not necessary to preserve chiral symmetry at the lattice scale, Wilson fermions could be ideal for this exploration. As we will see below, the simplest lattice field theory model with the above properties suffers from a sign problem in the traditional approach, but not in the fermion bag approach.

The model we consider contains an isospin doublet of Grassmann valued four-component spinor fields $\chib_{i,x},\chi_{i,x}$ $i=1,2$ on a four-dimensional space-time lattice, interacting with both an iso-scalar field $\sigma$ and a triplet of iso-vector fields $\vec{\pi}$. The action of the model is given by
\beq
S = S_b(\sigma,\vec{\pi}) + \sum_{i,j,x,y} \ \chib_{i,x} \ \tilde{M}_{i x , j y} \ \chi_{j,y}
\eeq 
where
\beq
\tilde{M}_{ix,jy} = \ D^{(w)}_{xy} \ \delta_{ij}
+ \ g\ \big(\sigma \delta_{ij}+ i \gamma_5 \vec{\tau}_{ij}\ \cdot\vec{\pi}_x\big)\ \delta_{xy}
\eeq
is the fermion matrix, $D^{(w)}$ is the Wilson Dirac operator,
\beq
D^{(w)}_{x,y} = m \ \delta_{xy} \ -\  \sum_{\mu} 
(r-\gamma_\mu) \delta_{x+\hat{\mu},y} +
(r+\gamma_\mu) \delta_{x-\hat{\mu},y} 
\label{wilson},
\eeq
$\vec{\tau}$ are the three Pauli matrices that act on the isospin space,  $\gamma_\mu$ and $\gamma_5$ are the five gamma matrices which we choose to represent by
\beqa
&\gamma_0 = \left(\begin{array}{cc} 1 & 0 \cr 0 & -1\end{array}\right),
\gamma_1 = \left(\begin{array}{cc} 0 & \sigma_1 \cr \sigma_1 & 0\end{array}\right),
\gamma_2 = \left(\begin{array}{cc} 0 & \sigma_2 \cr \sigma_2 & 0\end{array}\right),
&
\nonumber \\
&\gamma_3 = \left(\begin{array}{cc} 0 & \sigma_3 \cr \sigma_3 & 0\end{array}\right),
 \gamma_5 = \left(\begin{array}{cc} 0 & i \cr -i & 0\end{array}\right). &
\label{gammadef}
\eeqa
for later convenience and $S_b(\sigma, \vec{\pi)}$ is the $O(4)$ linear sigma model (\ref{o4model}). For convenience we have suppressed Dirac indices. Ignoring $D^{(w)}$ the action is invariant under an $O(4) \equiv SU(2) \times SU(2)$ symmetry, but including $D^{(w)}$ only preserves the vector $SU(2)$ symmetry.

The above model suffers from a sign problem in the traditional approach. To see this let us define $T = \gamma_2\gamma_5\tau_2$. Assuming $K$ is the complex conjugation operator it is easy to see that $[KT,\tilde{M}] = 0$ and $(KT)^2 = 1$. This proves that the determinant of $\tilde{M}$ is real but not necessarily positive \cite{Hands:2000ei}. For general background fields $(\sigma,\vec{\pi})$, $\mathrm{Det}\big(\tilde{M}\big)$ can indeed be negative. On the other hand, the above model belongs to the solvable class discussed in Sect.~\ref{sec3}. To see this, let us write the matrix $\tilde{M}$ in the  $2\times 2$ isospin space,
\beq
\tilde{M} = \left(\begin{array}{ccc} \ D_w + g\sigma + i g \gamma_5 \pi_3 & & 
i g \gamma_5(\pi_{1} - i\pi_{2}) \cr\cr
 i g \gamma_5(\pi_{1} + i\pi_{2}) & & D_w + g\sigma - i g \gamma_5 \pi_{3}
\end{array}\right),
\eeq
where we have separated the contribution from the four scalar fields. By redefining $\chib = \psib\gamma_5$, $\chi = i\tau_2 \psi$, $-i (\pi_{1} - i\pi_{2}) = \phi$, and $D(\sigma,\pi_3) =  \gamma_5 (D_w + g\sigma) + i g \pi_3$, we can rewrite the action as (\ref{newclass}) where,
\beq
M(\sigma,\pi_3) + g \ \Phi(\phi) = \left(\begin{array}{ccc} \  g \phi  && D (\sigma,\pi_3) \cr\cr
-D^\dagger(\sigma,\pi_3)  &&  g \phi^* \end{array}\right). 
\eeq
Since the bosonic action is solvable, the above model is free of sign problems in the fermion bag approach.

\section{More Solvable Models}
\label{sec6}

The Fermion bag approach is a general idea to perform Grassmann integrals. As long as one can define fermion bags carefully and identify symmetries (or other arguments) that ensure the positivity of fermion bag weights, sign problems are absent. The solutions to sign problems discussed in Sect.~\ref{sec3} considers the simplest class of models. As an example of a more complex situation, here we consider an exotic class of fermion models written in terms of two component Grassmann fields on a space-time lattice. They contain terms that naturally arise in the presence of spin-orbit couplings and can be interesting in condensed matter physics \cite{PhysRevLett.93.036403,PhysRevB.74.155426}.

Let $\psi_{\bfr,t,s},\psib_{\bfr,t,s},s=1,2$ be Grassmann fields that represent the two components of spin-half fermions hopping on a space-time lattice. We assume these fermions interact with two complex scalar fields that have their own dynamics. The lattice action is given by 
\beq
S = S_b(\varphi_1,\varphi_2) + S_f(\psib,\psi) + S_{\rm int}(\psib,\psi,\varphi_1,\varphi_2)
\label{exotic}
\eeq
 where
$S_b(\varphi_1,\varphi_2)$ is a solvable bosonic action whose form is not important,
\beqa
&& S_f(\psib,\psi) = - \sum_{\bfr,t,s} (\psib_{\bfr,{t+1},s} - \psib_{\bfr,t,s}) \psi_{\bfr,t,s} 
\nonumber \\
&& - \ \varepsilon \sum_{\bfr,\bfr',t,s,s'} 
\psib_{\bfr,t,s} \big(\delta_{s,s'} \alpha_{\bfr,\bfr'} - i  \vec{\sigma}_{s,s'} \cdot \vec{\beta}_{\bfr,\bfr'} \big)\psi_{\bfr',t}
\eeqa
is the free fermion action where $\alpha_{\bfr,\bfr'}$ is a real symmetric matrix of couplings, $\vec{\beta}_{\bfr,\bfr'}$ are three real anti-symmetric matrices of couplings, $\vec{\sigma}$ are Pauli matrices that act on the spin space and 
\beq
S_{\rm int} =  \varepsilon g 
\sum_{bfr, t}\ \big(\varphi^*_2(\bfr,t)\psib_{\bfr,t,2}\psib_{\bfr,t,1} + \varphi_1(\bfr,t)
\psi_{\bfr,t,1}\psi_{\bfr,t,2}\big)
\eeq
is the interaction term. For convenience we define the fermion matrix $M$ by expressing $S_f(\psib,\psi) = \psib M \psi$.

It is impossible to solve the sign problem in these class of models using the traditional approach. However, 
sign problems are absent in the fermion bag approach. The reason is that the $2k$-point correlation function
\beqa
&& \int \ [d\psib\ d\psi]\ \mathrm{e}^{-\psib\ M\ \psi} \ 
\big(\psi_{\bfr_1,t_1,1}\psi_{\bfr_1,t_1,2}...\psi_{\bfr_k,t_k,1}\psi_{\bfr_k,t_k,2}\big) 
\nonumber \\ 
&& 
\ \ \ \ \ \times\ \big(\psib_{\bfr'_1,t'_1,2}\psi_{\bfr'_1,t'_1,1}\ ...\ \psi_{\bfr'_k,t'_k,2}\psi_{\bfr'_k,t'_k,1}\big)
\label{kptcorr1} 
\eeqa
is positive and can be computed as a determinant exactly as in (\ref{fermcorr}). The proof involves the fact that $[KT,M] = 0$ where $K$ is the complex conjugation operator and $T = \sigma_2$. Since $(KT)^2 = -1$ it is possible to prove that $\mathrm{Det}(M) \geq 0$ \cite{Hands:2000ei}. Interestingly, the same proof is also applicable to any matrix obtained from $M$ by dropping rows or columns associated with both spin components at various lattice sites. Importantly, the sites where the rows are dropped need not be the same as the sites where the columns are dropped. This property ensures that all $2k$-point correlation functions (\ref{kptcorr1}) are positive. Thus, if the bosonic action is solvable then the above model is free of sign problems in the fermion bag approach.

In comparison to models in higher dimensions, there are many more models of interacting fermions in two space-time dimensions that are solvable using fermion bag ideas. For example, models like the one that was recently considered with an eight-fermion coupling \cite{PhysRevLett.109.250403}, can be handled with ease in the fermion bag approach. In one spatial dimension there are many fermion matrices that have real and positive determinants. Indeed since fermions in one spatial dimension can be mapped into hardcore bosons, fermion sign problems can be completely eliminated in the world line representation in many problems \cite{Evertz:2000rk}. In particular sign problems are absent at any density with both repulsive and attractive interactions. Quantum impurity problems in higher dimensions can also be mapped to one dimensional problems \cite{PhysRevB.71.201309}.  

\begin{table*}
\begin{center}
\caption{Comparison of all results obtained in the $N_f=2$ lattice Thirring model described by (\ref{ffmodels}) with $U_{\mcalb} = 0$. The results from the fermion bag approach are by far the most accurate to date.}
\label{tab:1} 
\begin{tabular}{ccccccc}
\hline\noalign{\smallskip}
Work & Range & Range & $U_c$ & $\nu$ & $\eta$ & $\eta_\psi$\\
& of $L$ & of $m$  &  &  &  & \\
\noalign{\smallskip}\hline\noalign{\smallskip}
mean field theory \cite{PhysRevLett.59.14} & - & 0 & 0.25 & 1 & 1 & 0 \\ 
\noalign{\smallskip}\noalign{\smallskip}
HMC results \cite{Debbio:1997dv} & 8-12 & 0.4-0.02 & 0.25(1) & 0.80(15) & 0.70(15) & - \\ 
\noalign{\smallskip}\noalign{\smallskip}
HMC results \cite{Barbour:1998yc} & 16-24 & 0.06-0.01 & 0.250(6) & 0.80(20) & 0.4(2) & - \\ 
\noalign{\smallskip}\noalign{\smallskip}
fermion bag Results \cite{PhysRevLett.108.140404} & 12-40 & 0 & 0.2608(2) & 0.85(1) & 0.65(1) & 0.37(1) \\ 
\noalign{\smallskip}\hline
\end{tabular}
% Or use
% \vspace*{5cm}  % with the correct table height
\end{center}
\end{table*}

\section{Computational Advantages}
\label{sec7}

In addition to solving sign problems, the fermion bag approach offers other computational advantages. For example, one of the bottlenecks in traditional methods is the inability to explore large system sizes even when interactions are weak or strong. This is because one always works with matrices that are as large as the system size.  On the other hand in the fermion bag approach the size of the matrices one deals with is related to the size of the fermion bags and not on the system size. The bag sizes are usually small at both weak and strong couplings, which makes the computational effort rather mild at these extremes. The real difficulties associated with large matrices are pushed into the intermediate coupling region. Even there one usually deals with bag sizes that are only a fraction of the physical volume, allowing one to explore large lattices with relative ease. Another advantage of the fermion bag approach is that recent developments in {\em worm} algorithms \cite{Prokof'ev:2001zz} can be used to sample the configuration space. These algorithms can help update topological properties of a configuration that are otherwise difficult to update. It was recently shown that a combination of worm algorithms and non-local updates eliminates all critical slowing down once the cost of computing determinant ratios is taken into account \cite{Chandrasekharan:2011vy}. Worm algorithms also help in measuring {\em off diagonal} observables that get contributions from configurations that do not contribute to the partition function.

\section{Results from a Four-Fermion Model}
\label{gnmodel}

The fermion bag approach was recently used to study a four-fermion model in three dimensions containing one flavor of lattice staggered fermions. Due to fermion doubling the model describes $N_f = 2$ flavors of four component Dirac fermions in the continuum. The action of the model is given by
\beqa
S &=& \sum_{x,y}\ \chib_x \big( D^{(s)}_{xy} + m \delta_{xy} \big) \chi_y - \ 
U_{\mcall}  \sum_{\langle xy \rangle \in \mcall } \ (\chib_x\chi_x) (\chib_y\chi_y)  
\nonumber \\
&& - \ U_\mcalb \sum_{\langle xy \rangle \in \mcalb } \ (\chib_x\chi_x) (\chib_y\chi_y)  
\label{ffmodels}
\eeqa
where the staggered Dirac matrix $D^{(s)}$ was defined in (\ref{3dstaggdirac}), $m$ is the fermion mass, $\langle xy \rangle$ labels two types of bonds : (1) link bonds $\mcall$ (between nearest neighbor sites) and (2) body diagonal bonds $\mcalb$ (between opposite sites of a body diagonal in cubes). When $m=0$, the model is invariant under an $SU(2)\times U(1)$ lattice symmetry. The $SU(2)$ symmetry arises when $\chi$ and $\chib$ on each site are treated as an $SU(2)$ doublet \cite{Catterall:2011ab}, while the $U(1)$ symmetry is the usual chiral symmetry of staggered fermions. The mass term breaks the $U(1)$ chiral symmetry.

Since the link coupling $U_{\mcall}$ looks like a current-current coupling, the model with $U_{\mcalb} = 0$ is referred to as the lattice Thirring model in the literature. It has been studied with traditional methods using the Hybrid Monte Carlo (HMC) algorithm \cite{DelDebbio:1995zc,Debbio:1997dv,Barbour:1998yc}. Due to difficulties in studying the chiral limit with the HMC algorithm, all previous Monte Carlo calculations were performed at non-zero fermion masses. The results were then extrapolated to the chiral limit and critical exponents were computed using the usual scaling analysis. Unfortunately the presence of two infrared scales, namely the box size and the fermion mass, makes such an analysis difficult. Recently, the fermion bag approach was also used to compute the same critical exponents. In contrast to the traditional approach, exactly massless fermions were studied and lattices as large as $40^3$ were explored \cite{PhysRevLett.108.140404}. Table \ref{tab:1} compares the results from various calculations performed so far. Although all results are consistent with each other, the fermion bag approach provides the most accurate results to date.

Thirring models in the continuum have a long history \cite{Parisi:1975im,Hikami:1976at}. They have been analyzed using various techniques including large $N_f$ \cite{Hands:1994kb}, Dyson-Schwinger equations \cite{Gomes:1990ed,Kondo:1995np,Hong:1993qk,Sugiura:1996xk}, and renormalization group flows \cite{Janssen:2012pq}. To compare lattice results with continuum results requires much care since the fixed point structure in the continuum is quite complex. A recent continuum calculation finds $\nu \approx 2.4$ and $\eta \approx 1.5$ \cite{Janssen:2012pq}, suggesting that the continuum Thirring universality is different from the universality of the lattice Thirring model. Is it possible that the lattice Thirring model actually flows to the fixed point of the $N_f=2$ Gross-Neveu (GN) model with a $U(1)$ chiral symmetry?

Lattice GN models also have a long history \cite{Karkkainen:1993ef,Hands:1995jq,PhysRevD.53.4616}. They can be constructed by introducing auxiliary scalar fields at centers of cubes. Each scalar field at the center of a cube couples to fermion mass terms at the corners \cite{Hands:1992be}.  In a theory with $U(1)$ chiral symmetry, integration over the auxiliary fields results in the model described by (\ref{ffmodels}) with non-zero values for both $U_{\mcall}$ and $U_{\mcalb}$. Comparing with the lattice Thirring model the only difference is that $U_{\mcalb} \neq 0$. Unfortunately, this non-zero coupling introduces sign problems in the traditional approach \cite{PhysRevD.85.091502}. Hence it would have been impossible to study an $N_f=2$ model previously. Yet, such a model with a $Z_2$ chiral symmetry seems to have been studied by the authors of \cite{Karkkainen:1993ef} and it was found that $\nu = 1.00(4)$ and $\eta = 0.754(8)$. Unfortunately, the sign problem was never addressed. These results however, are in excellent agreement with other theoretical analysis \cite{PhysRevLett.86.958}. It is possible to add conjugate fermion fields and eliminate the sign problem in the traditional approach, but at the expense of increasing $N_f$ by a factor of two. Such a modified model with a $U(1)$ chiral symmetry was recently studied on large lattices and it was found that $\nu = 1.03(4)$ and $\eta = 0.91(4)$ \cite{Christofi:2006zt}. 

The results for critical exponents obtained from lattice $GN$ models so far, also seem inconsistent with Tab. \ref{tab:1}. This discrepancy is some times used as evidence to argue that lattice $GN$ models and the lattice Thirring models belong to two different universality classes. If true, this leads to a disturbing conclusion that $U_{\mcalb} \neq 0$ could drive the transition into a different universality class although no lattice symmetries change! Although traditional methods suffer from a sign problem when $U_{\mcalb} \neq 0$, the fermion bag approach is free of sign problems even with $U_{\mcalb} \neq 0$ \cite{PhysRevD.85.091502} and can be used to compute the critical exponents. Not surprisingly one obtains the same critical exponents as in Tab. \ref{tab:1} suggesting that the universality class has not changed \cite{Li:2012dra}. Thus, we learn that the lattice Thirring model and the lattice GN model described by (\ref{ffmodels}) belong to the same universality class. But which one? And is there a theoretical framework to understand the critical exponents in Tab. \ref{tab:1}? 

Most results from lattice simulations in GN models are usually compared with large $N_f$ calculations. However, it was pointed out in \cite{Rosenstein:1993zf} that the series is poorly convergent for small values of $N_f$. It was claimed that a better approach would be to perform an $\epsilon = 4-d$ expansion. Results for $N_f=2$ up to order $\epsilon^2$ is given by \cite{Rosenstein:1993zf},
\beqa
& \nu = 0.5 + 0.23\epsilon + 0.12\epsilon^2,\ \ 
\eta = 0.57\epsilon + 0.10\epsilon^2,&
\nonumber \\
&\eta_\psi = 0.071 \epsilon - 0.009\epsilon^2.&
\eeqa
Substituting $\epsilon=1$ we obtain $\nu \approx 0.85$, $\eta \approx 0.67$ in excellent agreement with results of table \ref{tab:1}. However, $\eta_\psi \approx 0.062$ is completely off. Given the excellent agreement with the other two exponents, it is tempting to conclude that the lattice model described by (\ref{ffmodels}) belongs to the universality class of the continuum $N_f=2$ chiral $U(1)$ GN model close to the critical point. We think that by ignoring the sign problem in their calculations the authors of \cite{Karkkainen:1993ef} introduced conjugate fermions unintentionally and thus studied an $N_f=4$ model instead of an $N_f=2$ model. 

Further work is clearly needed to verify these claims. Given that $\eta_\psi$ is quite different from the theoretical prediction, it would be important to confirm it through an independent computation in a different model. Studies in GN models with a $Z_2$ chiral symmetry would be very useful. Finally, an independent calculation of other critical exponents could help in confirming hyper-scaling relations. All these are within reach of the fermion bag approach.

\section{Conclusions}

The fermion bag approach offers an alternative method to solve fermion sign problems. Instead of integrating over all Grassmann variables uniformly, they are grouped into bags such that Grassmann integrals can be performed separately inside each bag. By cleverly choosing the bags and the background bosonic fields many new sign problems can be solved. The new approach also teaches us that sign problems can be hidden in the choice of the bosonic fields and solutions may require the reformulation of the entire problem in new variables.

We have shown that many lattice field theories that are of physical interest but could not be studied until now, can be formulated without sign problems in the fermion bag approach. This creates new opportunities in the field of strongly correlated fermion systems some of which we have outlined in this work.

Three dimensional four-fermion models containing two flavors of Dirac fermions with a $U(1)$ chiral symmetry have been studied recently using the fermion bag approach. The critical exponents at the quantum critical point between the massless and massive phases could be computed accurately and the results 
($\nu = 0.85(1)$ and $\eta = 0.65(1)$) are in excellent agreement with $\epsilon$ expansion. On the other hand the anomalous dimension of the fermion field $\eta_\psi = 0.37(1)$ seems to disagree. 

\section{Acknowledgments}

I wish to thank H.-U.~Baranger, S.~Hands, D.~Kaplan, D.~Lee, A.~Li, B.~Rosenstein, and U.-J.~Wiese for discussions related to this work. This work was supported in part by the Department of Energy grant DE-FG02-05ER41368.

%
% BibTeX users please use
\bibliographystyle{science}
\bibliography{ref,graphene,cneft}
%
% Non-BibTeX users please use
%\begin{thebibliography}{}
%
% and use \bibitem to create references.
%
%\bibitem{RefJ}
% Format for Journal Reference
%Author, Journal \textbf{Volume}, (year) page numbers.
% Format for books
%\bibitem{RefB}
%Author, \textit{Book title} (Publisher, place year) page numbers
% etc
%\end{thebibliography}

\end{document}